\documentclass[conference]{IEEEtran}
\IEEEoverridecommandlockouts
\usepackage{cite}
\usepackage{amsmath,amssymb,amsfonts}
\usepackage{algorithmic}
\usepackage{graphicx}
\usepackage{textcomp}
\usepackage{xcolor}
\usepackage{subcaption}
\usepackage{hyperref}
\usepackage{booktabs}
\usepackage{pifont}
\newcommand{\cmark}{\ding{51}}  

\usepackage{multirow}
\usepackage{xcolor}
\usepackage{amssymb} 
\usepackage{booktabs}
\usepackage{tabularx}

\hypersetup{
  colorlinks=true,
  linkcolor=blue,     
  citecolor=green     
}
\def\BibTeX{{\rm B\kern-.05em{\sc i\kern-.025em b}\kern-.08em
    T\kern-.1667em\lower.7ex\hbox{E}\kern-.125emX}}
\begin{document}

\title{Precision Spatio-Temporal Feature Fusion for Robust Remote Sensing Change Detection\\
}

\author{
\IEEEauthorblockN{W.M.B.S.K. Wijenayake}
\IEEEauthorblockA{ 
\textit{University of Peradeniya}\\
Peradeniya, Sri Lanka \\
e19445@eng.pdn.ac.lk}
\and
\IEEEauthorblockN{R.M.A.M.B. Ratnayake}
\IEEEauthorblockA{ 
\textit{University of Peradeniya}\\
Peradeniya, Sri Lanka \\
e19328@eng.pdn.ac.lk}
\and
\IEEEauthorblockN{D.M.U.P. Sumanasekara}
\IEEEauthorblockA{ 
\textit{University of Peradeniya}\\
Peradeniya, Sri Lanka \\
e19391@eng.pdn.ac.lk}
\and
\IEEEauthorblockN{N.S. Wasalathilaka}
\IEEEauthorblockA{
\textit{University of Peradeniya}\\
Peradeniya, Sri Lanka \\
e20425@eng.pdn.ac.lk}
\and
\IEEEauthorblockN{M.Piratheepan}
\IEEEauthorblockA{ 
\textit{University of Peradeniya}\\
Peradeniya, Sri Lanka \\
e20293@ee.pdn.ac.lk}
\and
\IEEEauthorblockN{G.M.R.I. Godaliyadda}
\IEEEauthorblockA{
\textit{University of Peradeniya}\\
Peradeniya, Sri Lanka \\
roshang@eng.pdn.ac.lk}
\and
\IEEEauthorblockN{M.P.B. Ekanayake}
\IEEEauthorblockA{ 
\textit{University of Peradeniya}\\
Peradeniya, Sri Lanka \\
mpbe@eng.pdn.ac.lk }
\and
\IEEEauthorblockN{H.M.V.R. Herath}
\IEEEauthorblockA{ 
\textit{University of Peradeniya}\\
Peradeniya, Sri Lanka \\
vijitha@ee.pdn.ac.lk}
}

\maketitle

\begin{abstract}
Remote sensing change detection is vital for monitoring environmental and urban transformations but faces challenges like manual feature extraction and sensitivity to noise. Traditional methods and early deep learning models, such as convolutional neural networks (CNNs), struggle to capture long-range dependencies and global context essential for accurate change detection in complex scenes. While Transformer-based models mitigate these issues, their computational complexity limits their applicability in high-resolution remote sensing. Building upon ChangeMamba architecture, which leverages state space models for efficient global context modeling, this paper proposes precision fusion blocks to capture channel-wise temporal variations and per-pixel differences for fine-grained change detection. An enhanced decoder pipeline, incorporating lightweight channel reduction mechanisms, preserves local details with minimal computational cost. Additionally, an optimized loss function combining Cross Entropy, Dice and Lovasz objectives addresses class imbalance and boosts Intersection-over-Union (IoU). Evaluations on SYSU-CD, LEVIR-CD+, and WHU-CD datasets demonstrate superior precision, recall, F1 score, IoU, and overall accuracy compared to state-of-the-art methods, highlighting the approach’s robustness for remote sensing change detection. For complete transparency, the codes and pretrained models are accessible at \url{https://github.com/Buddhi19/MambaCD.git}
\end{abstract}

\begin{figure*}[t]
    \centering
    \includegraphics[width=1\linewidth]{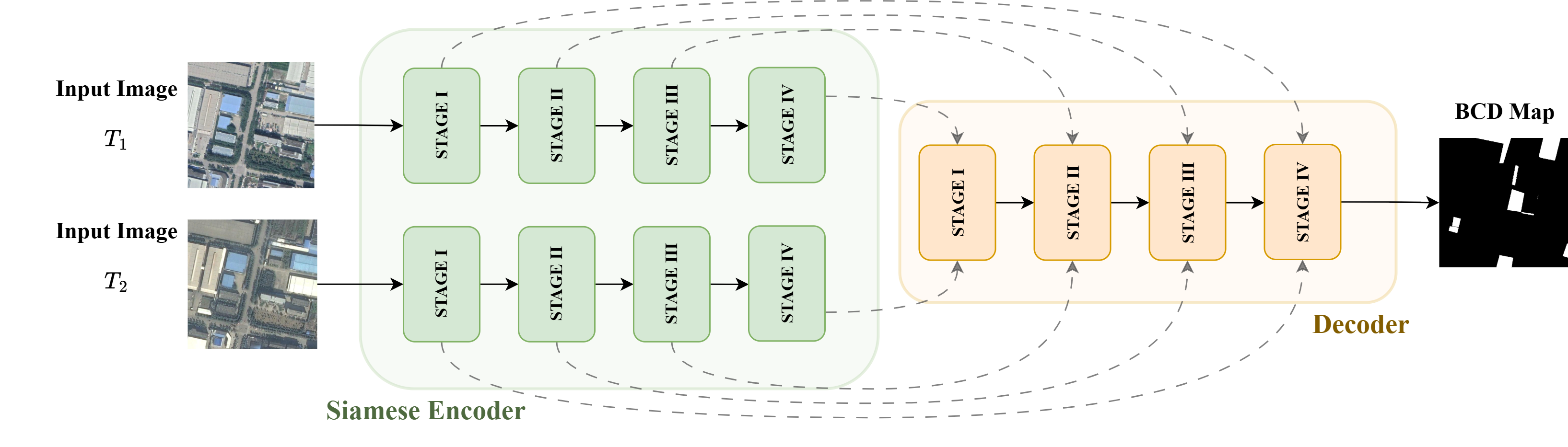}
    \caption{Overall Architecture of a Siamese Encoder-Decoder Framework for Binary Change Detection}
    \label{fig:overall_architecture}
\end{figure*}
\section{Introduction}
Change detection (CD) is a fundamental technique for identifying temporal alterations in specific objects or areas using multi-temporal satellite imagery. In remote sensing applications, CD leverages time-series satellite data to detect and analyze surface changes on Earth, with applications spanning agriculture, urban environmental monitoring, disaster management, urban expansion analysis, and natural disaster damage assessment including fires and floods \cite{C_Cheng_2024_5043}.

Recent advances in satellite technology have revolutionized remote sensing capabilities through improved spatial, temporal, and spectral resolution, enabling detailed monitoring of Earth's surface changes \cite{S_Wang_2023}. The deployment of low-earth orbit satellites with increased revisit frequencies has enhanced real-time monitoring capabilities and data availability \cite{C_L._2024}. 

However, traditional remote sensing and change detection methods\cite{A_Bruzzone_2000} face significant limitations, including reliance on manual pattern identification by human experts, making the process time-consuming and highly dependent on domain knowledge\cite{A_Jiang_2022}. Additionally, conventional pixel-based approaches are particularly sensitive to atmospheric variations, noise, illumination changes, and registration errors\cite{C_Cheng_2024}.

The emergence of deep learning has revolutionized change detection in remote sensing imagery, addressing fundamental limitations of traditional methods. Rather than relying on handcrafted features, convolutional neural networks (CNNs) automatically learn hierarchical representations from low-level textures to high-level semantics. Early CNN architectures such as FC-EF and FC-Siam-diff established the foundation for deep learning-based change detection by learning discriminative features from bi-temporal data \cite{F_Caye_2018}. Siamese networks became widely adopted for consistent feature extraction across temporal sequences \cite{S_Fang_2022_f6ff}. However, CNNs are inherently constrained by their local receptive fields, limiting their ability to capture long-range dependencies and global context crucial for accurately identifying changes across large or complex scenes.

Vision Transformers address CNN limitations in modeling long-range dependencies through self-attention mechanisms that capture global spatial relationships, making them highly effective for change detection tasks\cite{A_Beyer_2020}. Transformer-based methods such as ChangeFormer\cite{A_Gedara_2022}, tokenize image patches and apply multi-head self-attention to understand contextual relationships between changes and the broader scene, offering enhanced interpretability through attention visualization. However, the quadratic complexity of self-attention in Transformers presents computational challenges for high-resolution remote sensing applications\cite{keles2022computationalcomplexityselfattention}
.

The Mamba architecture\cite{gu2024mambalineartimesequencemodeling} represents a significant advancement in balancing global context modeling with computational efficiency through State Space Models (SSMs). Unlike attention-based Transformers with high computational requirements, Mamba employs a selective scan algorithm with linear computational complexity, enabling efficient processing of high-resolution remote sensing images. \textit{ChangeMamba} adapts this backbone to bi-temporal remote sensing inputs, achieving promising results for change detection tasks \cite{C_Chen_2024}. However, ChangeMamba may not fully exploit fine-grained temporal changes, such as channel-wise differences, and its decoder might discard valuable local context during feature fusion. Furthermore, optimizing for key metrics like Intersection over Union (IoU) in the presence of class imbalance remains a challenge.

Building upon ChangeMamba, we propose the following key contributions to address these gaps,
\begin{enumerate}
  \item \textbf{Precision fusion blocks} that introduce \emph{channel-wise temporal cross modeling} and an explicit \emph{difference module} to capture per-channel and per-pixel changes more effectively.
  \item \textbf{Enhanced decoder pipeline} that replaces the single $1{\times}1$ bottleneck with lightweight depthwise-separable convolutions followed by a Convolutional Block Attention Module, preserving local detail with minimal computational overhead.
  \item \textbf{Improved optimization strategy} that incorporates Dice loss with the cross-entropy + Lov\'{a}sz objective, mitigating class imbalance and directly promoting higher IoU.
\end{enumerate}

\section{Methodology}

\begin{figure*}[t]
    \centering
    \includegraphics[width=1\linewidth]{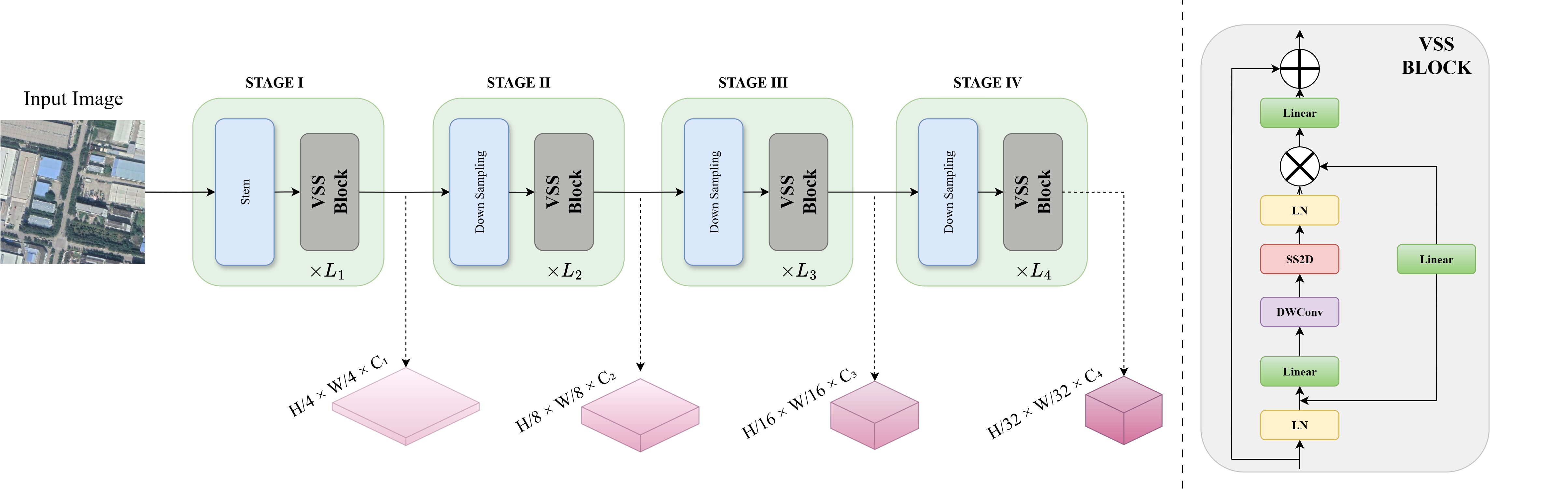}
    \caption{Structure of the dual-stream VMamba encoder}
    \label{fig:encoder}
\end{figure*}
\subsection{ChangeMamba}\label{sec:changemamba}

As depicted in Figure~\ref{fig:overall_architecture}, the ChangeMamba architecture, for BCD task~\cite{C_Chen_2024}, integrates two core components, a Visual Mamba (VMamba) encoder, inspired by State Space Models (SSM)~\cite{zhu2024visionmambaefficientvisual}, and a change decoder. The VMamba encoder, illustrated in Figure~\ref{fig:encoder}, processes input images from two distinct time steps, pre-event (\(T_1\)) and post-event (\(T_2\)), to generate hierarchical feature representations that capture global spatial contextual information. These features, denoted as \( F^{T_1}_{i,j} \) and \( F^{T_2}_{i,j} \in \mathbb{R}^{H_i \times W_i \times C_i} \) for stage \( i \) and block \( j \), are subsequently fed into the change decoder, shown in Figure~\ref{fig:decoder}. The architecture offers three variants named Tiny, Small, and Base, distinguished by the number of Visual State Space (VSS) blocks and feature channels, with the Base variant typically outperforming other variants, making it the preferred choice for this study with $C_1=128, C_2 = 256, C_3 = 526, C_4 = 1024, L_1 = 2, L_2 = 2, L_3 = 15$ and $ L_4 = 2$.

To effectively integrate features from \(T_1\) and \(T_2\), ChangeMamba employs several spatio-temporal relationship modeling mechanisms within the STSS blocks of the change decoder as seen in \ref{fig:decoder}\textcolor{blue}{a}. These mechanisms, detailed below as presented in~\cite{C_Chen_2024}, form the backbone of the architecture’s change detection capabilities.

\subsubsection{Spatio-temporal Sequential Modeling}\label{modeling1}
Spatio-temporal sequential modeling organizes feature tokens from pre-event and post-event images into a temporally ordered sequence. For each stage \( i \) and block \( j \), the feature tokens from the pre-event feature map \( F^{T_1}_{i,j} \) are arranged first, followed by those from the post-event feature map \( F^{T_2}_{i,j} \). This forms a single sequence that enables the model to process the pre-event state before the post-event state, expressed as,

\begin{equation}
F^\text{seq}_{i,j} = \left[ F^{T_1}_{i,j}(1), \dots, F^{T_1}_{i,j}(N), F^{T_2}_{i,j}(1), \dots, F^{T_2}_{i,j}(N) \right]
\label{eq:seq_model}
\end{equation}
where \( N \) represents the number of spatial locations in the feature map after downsampling at stage \( i \) and block \( j \), given by $N = \frac{HW}{2^{1+j}}$.

The resulting sequence \( F^\text{seq}_{i,j} \in \mathbb{R}^{H_i \times 2W_i \times C_i} \) captures both spatial and temporal features (see Figure \ref{fig:decoder}\textcolor{blue}{a}).

\subsubsection{Spatio-temporal Cross Modeling}
Spatio-temporal Cross Modeling interleaves the feature tokens \(F^{T_1}_{i,j}\) and \(F^{T_2}_{i,j}\) to facilitate direct interaction between corresponding spatial locations across the two time steps. For each spatial position, the tokens from \(T_1\) and \(T_2\) are alternated in the sequence, enabling the model to compare features at the same location across different times. This can be represented as,

\begin{equation}
F^\text{crs}_{{i,j}} = \left[F^{T_1}_{i,j}(1), F^{T_2}_{i,j}(1), \dots, F^{T_1}_{i,j}(N), F^{T_2}_{i,j}(N)\right]
\label{eq:crs_model}
\end{equation}

The resulting sequence \( F^\text{crs}_{i,j} \in \mathbb{R}^{H_i \times 2W_i \times C_i} \) captures the interleaved spatio-temporal features(Figure \ref{fig:decoder}\textcolor{blue}{a}).

\subsubsection{Spatio-temporal Parallel Modeling}
Spatio-temporal parallel modeling concatenates feature tokens from pre-event (\( F^{T_1}_{i,j} \)) and post-event (\( F^{T_2}_{i,j} \)) images along the channel dimension, enabling simultaneous processing of both temporal states at each spatial location. This is expressed as,
\begin{equation}
F^\text{pra}_{i,j} = F^{T_1}_{i,j} \mathbin{\mathord{\text{\textcircled{c}}}} F^{T_2}_{i,j}
\label{eq:pra_model}
\end{equation}

where \( \mathbin{\mathord{\text{\textcircled{c}}}} \) denotes concatenation along the channel dimension. The resulting feature map \( F^\text{pra}_{i,j} \in \mathbb{R}^{H_i \times W_i \times 2C_i} \)(Figure \ref{fig:decoder}\textcolor{blue}{a}).

\subsection{Precision Fusion Blocks}\label{modeling2}

To enhance feature fusion, we introduce two additional modeling mechanisms, described below.

\subsubsection{\textbf{Channel-wise Temporal Cross Modeling}}\label{sec:chn}
\begin{figure*}[ht]
    \makebox[\textwidth][c]{\includegraphics[width=1.1\textwidth]{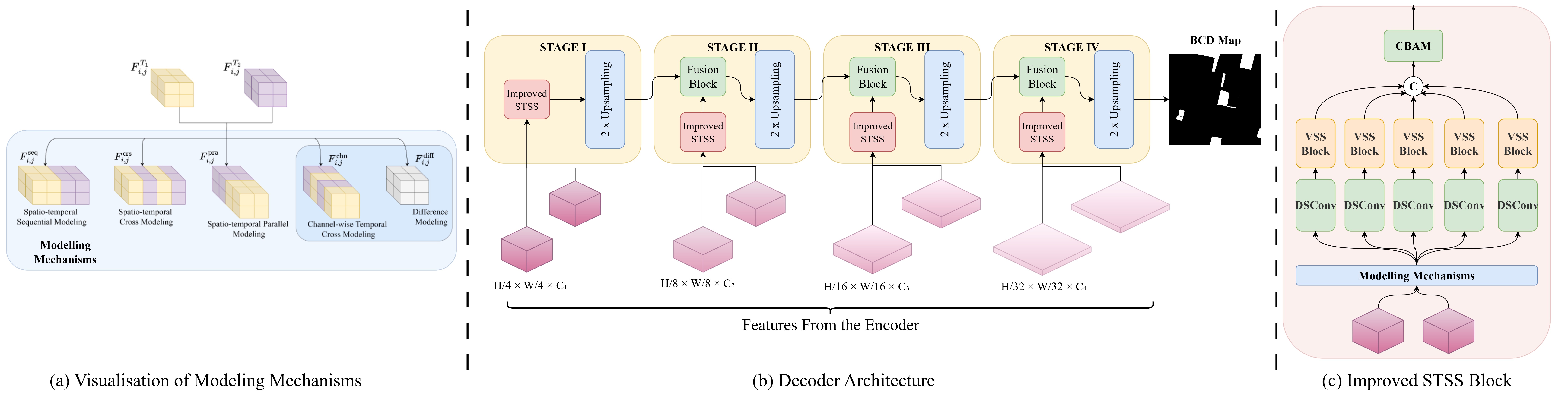}}
    \caption{Enhanced decoder with Improved STSS Block}
    \label{fig:decoder}
\end{figure*}

Channel-wise temporal cross modeling interleaves pre-event (\( F^{T_1}_{i,j} \)) and post-event (\( F^{T_2}_{i,j} \)) feature tensors along the channel dimension to capture fine-grained temporal differences. For feature tensors \( F^{T_1}_{i,j}, F^{T_2}_{i,j} \in \mathbb{R}^{H_i \times W_i \times C_i} \), the resulting tensor \( F^{\text{chn}}_{i,j} \in \mathbb{R}^{H_i \times W_i \times 2C_i} \) is constructed as,
\begin{equation}
\small
\begin{aligned}
& F^{\text{chn}}_{i,j}(h,w,:) =\\
&\left[ F^{T_1}_{i,j}(h,w,1), F^{T_2}_{i,j}(h,w,1),.., F^{T_1}_{i,j}(h,w,C_i), F^{T_2}_{i,j}(h,w,C_i) \right]
\label{eq:chn_array}
\end{aligned}
\end{equation}

where \( h \) and \( w \) denote spatial coordinates, and the colon ($:$) represents all channels. This interleaving creates a zipped pattern, where each pre-event channel is immediately followed by its post-event counterpart (see Figure \ref{fig:decoder}\textcolor{blue}{a}). 

\subsubsection{\textbf{Difference Modeling}}
Unlike previous strategies that implicitly encode temporal cues through ordering or interleaving, difference modeling explicitly captures temporal changes by computing the per-pixel feature residual between pre-event and post-event tensors. For encoder outputs \( F^{T_1}_{i,j}, F^{T_2}_{i,j} \in \mathbb{R}^{H_i \times W_i \times C_i} \), we define the difference tensor \( F^{\text{diff}}_{i,j} \in \mathbb{R}^{H_i \times W_i \times C_i} \) as,
\begin{equation}
F^{\text{diff}}_{i,j} = \left| F^{T_2}_{i,j} - F^{T_1}_{i,j} \right|
\label{eq:signed_diff}
\end{equation}
where \( \left| \cdot \right| \) denotes the absolute value, emphasizing the magnitude of temporal changes at each spatial location.

\subsection{Enhanced Channel Reduction (ECR)}\label{sec:ds_cbam}
In the original ChangeMamba decoder, each modelling mechanism output is projected to 128 channels before processing by VSS blocks (Figure \ref{fig:decoder}\textcolor{blue}{c}), and after concatenating VSS outputs, a \( 1\times1 \) convolution further reduces the channels to 128. This simple projection may discard valuable local context. To address this, we introduce two enhancements,

\begin{enumerate}
  \item \textbf{Depthwise Separable Convolution (DSConv).}  
    A lightweight depthwise separable convolution\cite{chollet2017xceptiondeeplearningdepthwise}
 is applied to each modeling mechanism output ($F^{\text{mech}}_{i,j}$), to efficiently mix spatial and channel information as,
    \begin{equation}
    F^{\text{DS-mech}}_{i,j} = \text{PW}^{128}\bigl(\text{DW}^{3\times3}(F^{\text{mech}}_{i,j})\bigr)
    \label{eq:dsconv}
    \end{equation}
    where \( \text{DW}^{3\times3} \) denotes a depthwise convolution with a \( 3\times3 \) spatial filter per channel, and \( \text{PW}^{128} \) is a pointwise convolution projecting the output to 128 channels. The resulting outputs ($F^{\text{DS-seq}}_{i,j}$, $F^{\text{DS-crs}}_{i,j}$, $F^{\text{DS-pra}}_{i,j}$, $F^{\text{DS-chn}}_{i,j}$, $F^{\text{DS-diff}}_{i,j} \;\in\mathbb{R}^{H_i \times W_i \times 128}$) are each processed by a VSS block. These outputs are then concatenated along the channel dimension to form $P_i\in\mathbb{R}^{H_i \times W_i \times 640}$.
    
  \item \textbf{CBAM Refinement}  
    After concatenating VSS outputs at each decoder stage, a Convolutional Block Attention Module (CBAM) \cite{woo2018cbamconvolutionalblockattention}
refines the fused feature map \( P_i \) as,

    \begin{equation}
    \small
    \begin{aligned}
    M_c &= \sigma \bigl( \text{MLP} ( \text{AvgPool}(P_i) ) + \text{MLP} ( \text{MaxPool}(P_i) ) \bigr) \\
    M_s &= \sigma \bigl( \text{Conv} ( \text{AvgPool}_c(P_i) \mathbin{\mathord{\text{\textcircled{c}}}} \text{MaxPool}_c(P_i) ) \bigr) \\
    \widetilde{P}_i &= M_s \odot ( M_c \odot P_i )
    \end{aligned}
    \end{equation}
    where \( \sigma \) is the sigmoid activation, \( \odot \) denotes element-wise multiplication, and \( \mathbin{\mathord{\text{\textcircled{c}}}} \) indicates channel-wise concatenation. The channel attention mask \( M_c \) emphasizes informative feature maps, while the spatial attention mask \( M_s \) highlights regions relevant to changes. The refined feature map \( \widetilde{P}_i \in \mathbb{R}^{H_i \times W_i \times 128} \) is passed to the next decoder stage, enhancing focus on salient changes.
\end{enumerate}

We update the original decoder architecture as seen in \ref{fig:decoder}\textcolor{blue}{b}, combining features from the pre-event and post-event encoders to generate a change map. It operates across multiple stages, each aligned with a resolution level from the encoder. At each stage, Improved STSS blocks process the encoder features, modeling spatio-temporal relationships through mechanisms as explained in \ref{modeling1} and \ref{modeling2} , integrating multi-scale features into a unified representation, ultimately producing a change map highlighting differences between the input images.

\subsection{Loss Function}

Our model is trained using a combination of three loss functions. In addition to Cross Entropy Loss and Lov\'{a}sz Loss\cite{lovaaaas}
 introduced in\cite{C_Chen_2024}  , we use Dice Loss\cite{Sudre_2017}
. These losses are chosen to optimize the model's performance addressing challenges such as class imbalance and directly optimizing key evaluation metrics like Intersection over Union (IoU).

The Cross Entropy Loss is a standard loss function for binary classification tasks. It is defined as,
\begin{equation}
\mathcal{L}_{CE} = -\frac{1}{N} \sum_{i=1}^{N} \left[ y_i \log(\hat{y}_i) + (1 - y_i) \log(1 - \hat{y}_i) \right]
\end{equation}
where \( N \) is the total number of pixels, \( y_i \) is the ground truth label for pixel \( i \) (1 for change, 0 for no change), and \( \hat{y}_i \) is the predicted probability of change for pixel \( i \).

The Lov\'{a}sz Loss\cite{lovaaaas}
 is a surrogate loss function that directly optimizes the Jaccard index (IoU)\cite{costa2021generalizationsjaccardindex}
. It is particularly effective for segmentation tasks with class imbalance. For binary segmentation, the Lov\'{a}sz Loss is computed based on the sorted errors between the predicted probabilities and the ground truth labels, providing a tractable way to optimize the IoU metric.

\subsubsection{\textbf{Dice Loss}}
The Dice Loss is based on the Dice coefficient, which measures the overlap between the predicted and ground truth change regions. It is defined as,
\begin{equation}
\mathcal{L}_{dice} = 1 - \frac{2 \sum_{i=1}^{N} y_i \hat{y}_i}{\sum_{i=1}^{N} y_i + \sum_{i=1}^{N} \hat{y}_i}
\end{equation}
where \( y_i \) and \( \hat{y}_i \) are as defined above. This loss function encourages the model to maximize the overlap between the predicted change areas and the actual change areas, thus improving the precision of change detection.

Hence the total loss function used for training is,

\begin{equation}
\mathcal{L}_{\text{total}} = \mathcal{L}_{CE} + 0.5 \mathcal{L}_{\text{Lovasz}} + 0.35 \mathcal{L}_{\text{dice}}
\end{equation}

where the coefficients were determined through experimentation, allowing the model to benefit from the strengths of each individual loss function.
\section{Experiments} \subsection{Datasets} \subsubsection*{1) SYSU-CD~\cite{A_Shi_2022}} SYSU-CD includes 20,000 pairs of 0.5m resolution aerial images from Hong Kong (2007--2014), capturing urban and coastal changes like building developments and sea reclamation. It uses a 6:2:2 split with 256×256 pixel patches.

\subsubsection*{2) LEVIR-CD+~\cite{A_Chen_2020}} LEVIR-CD+ comprises 985 pairs of 0.5m resolution aerial images (1024×1024 pixels) focusing on building-level changes over 5--14 years, covering construction and demolition of various building types.

\subsubsection*{3) WHU-CD~\cite{F_Ji_2019}} WHU-CD contains 0.3m resolution aerial images from Christchurch, New Zealand (2012 and 2016), targeting large-scale building changes. Images are divided into 256×256 patches, with 6,096 for training, 762 for validation, and 762 for testing.

\subsection{Experimental Setup} \subsubsection{Evaluation Metrics} We evaluate performance using recall (Rec), precision (Pre), overall accuracy (OA), F1 score, Intersection over Union (IoU), and Kappa coefficient (KC). Higher values indicate better performance, with definitions in~\cite{A_Zhu_2024}.
\begin{figure*}[t]
    \makebox[\textwidth][c]{%
        \begin{subfigure}[t]{0.353\textwidth}
            \centering
            \includegraphics[width=\linewidth]{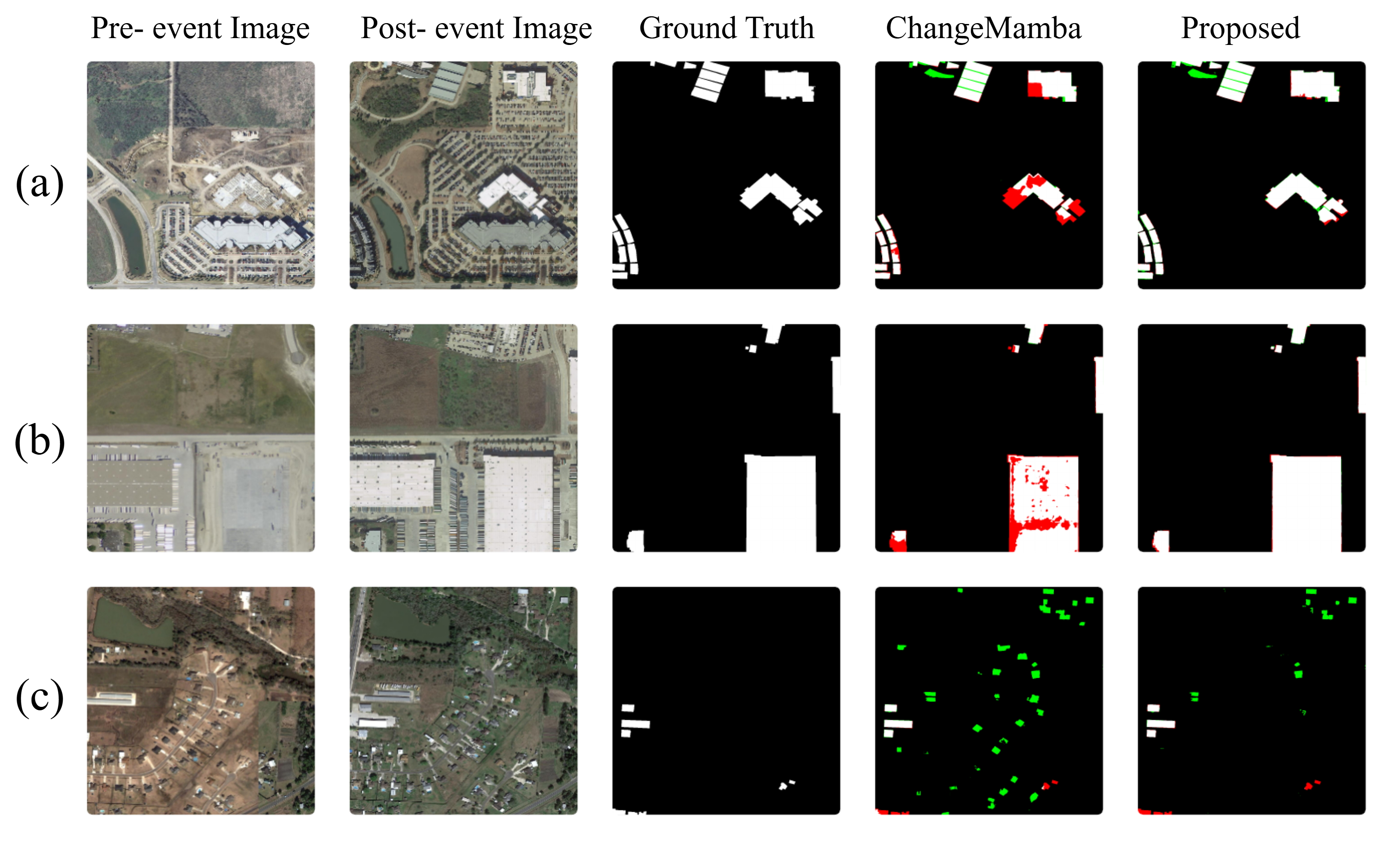}
            \caption{LEVIR-CD+ test set}
            \label{fig:levir}
        \end{subfigure}%
        \begin{subfigure}[t]{0.353\textwidth}
            \centering
            \includegraphics[width=\linewidth]{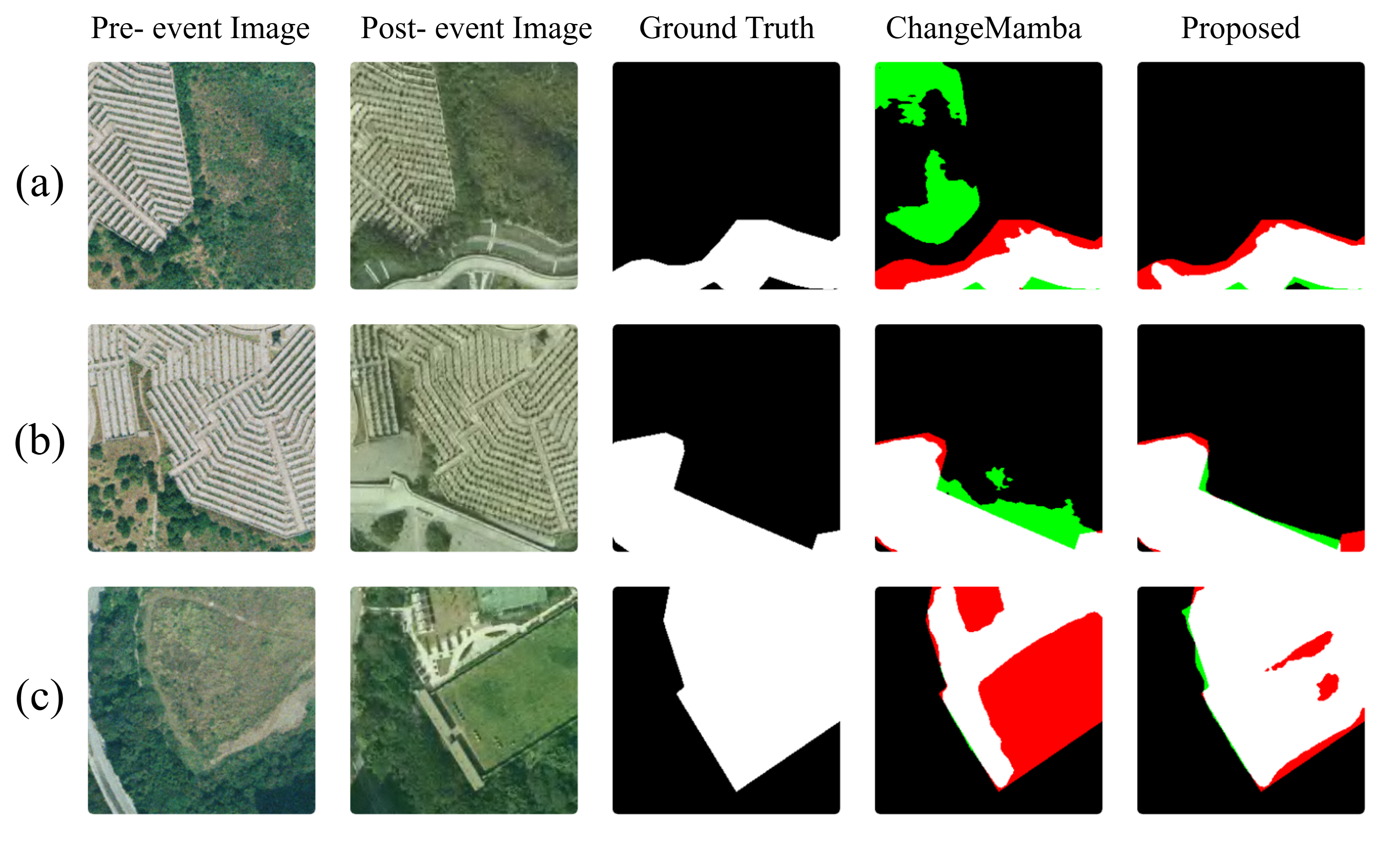}
            \caption{SYSU-CD test set}
            \label{fig:sysu}
        \end{subfigure}%
        \begin{subfigure}[t]{0.353\textwidth}
            \centering
            \includegraphics[width=\linewidth]{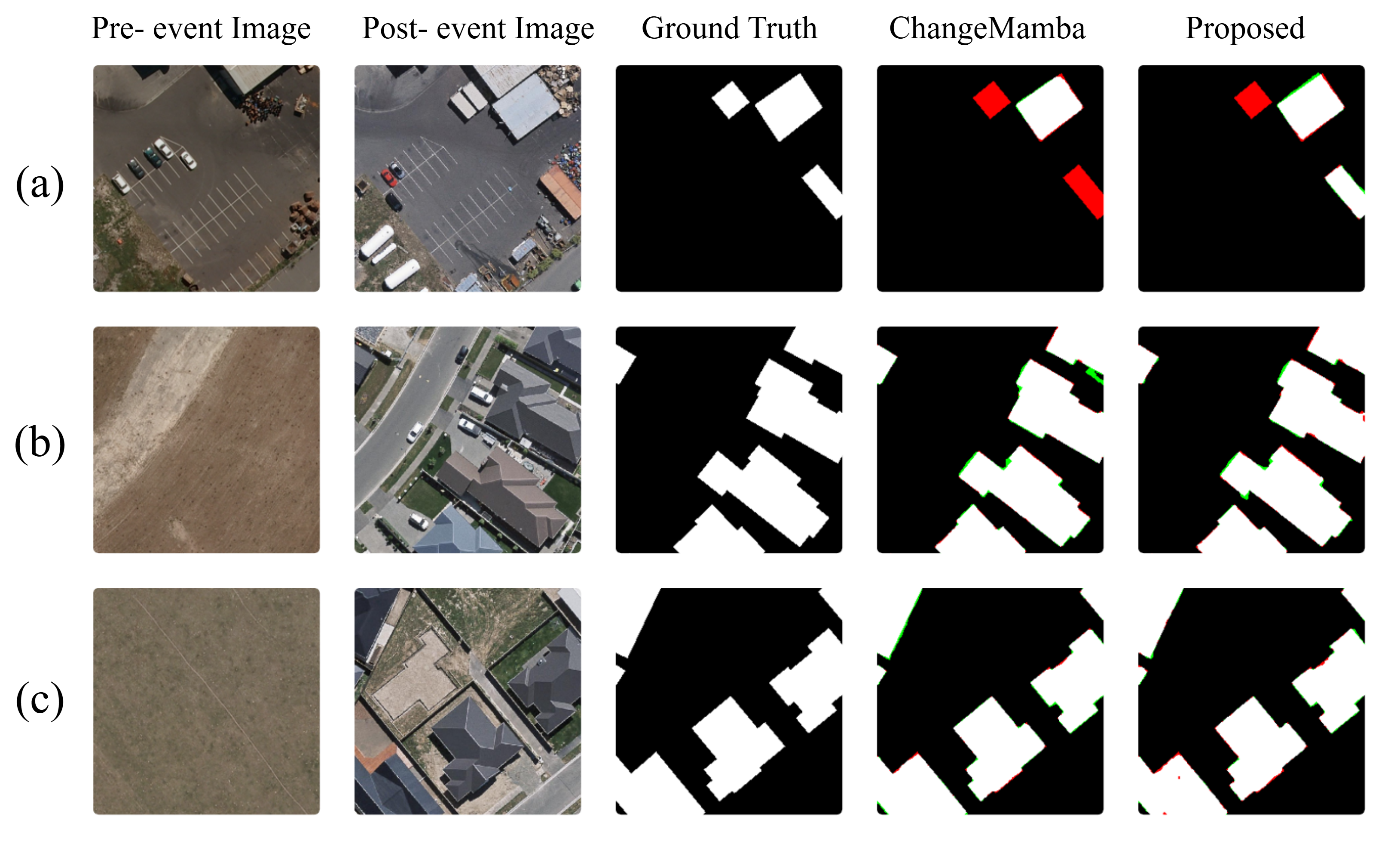}
            \caption{WHU-CD test set}
            \label{fig:whu}
        \end{subfigure}%
    }
    \caption{Qualitative visualization of change detection results on the LEVIR-CD+, SYSU-CD, and WHU-CD test sets. White represents true positives, black represents true negatives, \textcolor{red}{red} indicates false positives, and \textcolor{green}{green} indicates false negatives.}
    \label{fig:combined_results}
\end{figure*}

\subsubsection{Implementation Details}
Image pairs and their corresponding ground truth labels are cropped into 256 $\times$ 256 pixel patches for network input during training. For testing, the trained networks are applied to the original full-resolution images in the test set. The network is optimized using the AdamW optimizer~\cite{adam}
 with a learning rate of $1 \times 10^{-4}$ and a weight decay of $5 \times 10^{-3}$. The batch size is set to 4. Training is conducted for 50,000 training iterations.


\begin{table*}
\centering
\footnotesize
\caption{Comparison results across datasets for different change detection (CD) models. All results are in percentage (\%). Highest value for each metric is highlighted in \textcolor{red}{red}, the second highest in \textcolor{blue}{blue}, and the third highest in \textcolor{green}{green}.}
\label{tab:cd_comparison}
\begin{tabular}{p{0.1cm}l
  ccccc
  ccccc
  ccccc}
\toprule
 &  & \multicolumn{5}{c}{WHU-CD} & \multicolumn{5}{c}{LEVIR-CD} & \multicolumn{5}{c}{SYSU} \\
\cmidrule(lr){3-7} \cmidrule(lr){8-12} \cmidrule(l){13-17}
 & Model 
  & Pre & Rec & F1 & IoU & OA 
  & Pre & Rec & F1 & IoU & OA 
  & Pre & Rec & F1 & IoU & OA \\
\midrule
\multirow{5}{*}{$\mathcal{C}$}
  & FC-Siam-Conc \cite{F_Caye_2018}      & 84.02 & 87.72 & 85.83 & 75.18 & 98.94 
                       & 78.39 & 78.49 & 78.44 & 64.53 & 98.24 
                       & 73.67 & 76.75 & 75.18 & 60.23 & 88.05 \\
  & SiamCRNN-101 \cite{C_Chen_2020}      & 87.79 & 90.45 & 89.10 & 80.34 & 99.19 
                       & 85.56 & 80.96 & 83.20 & 71.23 & 98.67 
                       & 80.40 & \textcolor{green}{80.48} & 80.44 & 67.28 & 90.77 \\
  & SNUNet \cite{S_Fang_2022}            & 88.04 & 87.36 & 87.70 & 78.09 & 99.10 
                       & 71.07 & 78.73 & 74.70 & 59.62 & 97.83 
                       & 74.09 & 72.21 & 73.14 & 57.66 & 87.49 \\
  & DSIFN\cite{A_Zhang_2020}              & \textcolor{red}{97.46} & 83.45 & 89.91 & 81.67 & 99.31 
                       & 83.78 & 84.36 & 84.07 & 72.52 & 98.70 
                       & 75.83 & \textcolor{blue}{82.02} & 78.80 & 65.02 & 89.59 \\
  & CGNet  \cite{C_Han_2023}            & 94.47 & 90.79 & 92.59 & 86.21 & 99.48 
                       & 81.46 & \textcolor{green}{86.02} & 83.68 & 71.94 & 98.63 
                       & \textcolor{red}{86.37} & 74.37 & 79.92 & 66.55 & 91.19 \\
\midrule
\multirow{4}{*}{$\mathcal{T}$}
  & ChangeFormerV6 \cite{A_Gedara_2022}    & 85.49 & 81.90 & 83.66 & 71.91 & 98.83 
                       & 67.66 & 78.57 & 72.71 & 57.12 & 97.60 
                       & 81.70 & 72.38 & 76.76 & 62.29 & 89.67 \\
  & TransUNetCD \cite{T_Li_2022}       & 85.48 & 90.50 & 87.79 & 78.44 & 99.09 
                       & 83.08 & 84.18 & 83.63 & 71.86 & 98.66 
                       & 82.59 & 77.73 & 80.09 & 66.79 & 90.88 \\
  & SwinSUNet \cite{S_Zhang_2022}         & 94.08 & \textcolor{green}{92.03} & 93.04 & 87.00 & 99.50 
                       & 85.34 & 85.85 & \textcolor{green}{85.60} & \textcolor{green}{74.82} & \textcolor{green}{98.92} 
                       & 83.50 & 79.75 & \textcolor{green}{81.58} & \textcolor{green}{68.89} & \textcolor{green}{91.51} \\
  & CTDFormer \cite{R_Zhang_2023}        & 92.23 & 85.37 & 88.67 & 79.65 & 99.20 
                       & 80.58 & 80.03 & 80.30 & 67.09 & 98.40 
                       & 80.80 & 75.53 & 78.08 & 64.04 & 90.00 \\
\midrule
\multirow{2}{*}{$\mathcal{M}$}
  & MambaBCD-Base \cite{C_Chen_2024}     & \textcolor{blue}{96.18} & \textcolor{blue}{92.23} & \textcolor{blue}{94.19} & \textcolor{blue}{89.02} & \textcolor{blue}{99.58} 
                       & \textcolor{blue}{89.24} & \textcolor{blue}{87.57} & \textcolor{blue}{88.39} & \textcolor{blue}{79.20} & \textcolor{blue}{99.06} 
                       & \textcolor{green}{86.11} & 80.31 & \textcolor{blue}{83.11} & \textcolor{blue}{71.10} & \textcolor{blue}{92.30} \\
  & \textbf{Proposed}  & \textcolor{green}{95.95} & \textcolor{red}{93.50} & \textcolor{red}{94.71} & \textcolor{red}{89.95} & \textcolor{red}{99.60} 
                       & \textcolor{red}{91.38} & \textcolor{red}{90.43} & \textcolor{red}{90.90} & \textcolor{red}{83.32} & \textcolor{red}{99.26} 
                       & \textcolor{blue}{86.14} & \textcolor{red}{85.35} & \textcolor{red}{85.74} & \textcolor{red}{75.04} & \textcolor{red}{93.30} \\
\bottomrule
\end{tabular}
\end{table*}

\begin{table}[t]
\centering
\footnotesize
\caption{Ablation study on the SYSU-CD dataset. All results are in \%.}
\label{tab:ablation}
\begin{tabular}{p{0.5cm}p{0.5cm}p{0.5cm}p{0.5cm}ccccc}
\toprule
\multicolumn{4}{c}{Components} & \multicolumn{5}{c}{Metrics} \\
\cmidrule(lr){1-4} \cmidrule(l){5-9}
$F^{\text{diff}}_{i,j}$ & $F^{\text{chn}}_{i,j}$ & $L_{Dice}$ & ECR & Pre & Rec & F1 & IoU & OA \\
\midrule
   & \cmark & \cmark & \cmark &  88.96   &  80.94   &  84.76   &  73.56   &   93.14  \\
\cmark &   & \cmark & \cmark &   88.47  &   80.40  &   84.24  &   72.78  &  92.91   \\
\cmark & \cmark &   & \cmark &  86.19   &  84.13   &  85.15   &   74.13  &  93.08   \\
 \cmark & \cmark & \cmark &   &    88.00 &  82.85   &  85.35& 74.40    &  93.10   \\
\cmark & \cmark & \cmark & \cmark &86.14 &85.35 &85.74 &75.04 &93.30    \\
\bottomrule
\end{tabular}
\end{table}

\section{Results and Discussion}
    The proposed algorithm was evaluated for the chosen datasets and compared against the State-of-the-Art (SOTA) algorithms whose results are summarized in Table \ref{tab:cd_comparison}. Furthermore, exhaustive ablation studies were performed to investigate the impact of the key novelties in the proposed algorithm on performance. 

    Considering Table \ref{tab:cd_comparison}, it is clear that the proposed algorithm outperforms the current SOTA algorithms in most metrics across all the datasets. The only exception is Precision, where the proposed algorithm is second best for SYSU and third best for the WHU-CD dataset. Compared with the difference between the performance of other algorithms, it is clear that the improvement of the proposed method is significant.

    The competitive performance of the proposed method is further cemented by the visual analysis of the Qualitative Comparison with the ChangeMamba Algorithm presented in Figure \ref{fig:levir}, Figure \ref{fig:sysu}, and Figure \ref{fig:whu}. Observing Figure \ref{fig:levir}, it is clear that the proposed algorithm's output more closely adheres to the Ground Truth. Similarly, observing Figure \ref{fig:sysu} for the SYSU-CD dataset, the proposed algorithm has very few false negative regions, while false positive regions are much diminished when compared with the ChangeMamba Algorithm. Similar, though less pronounced, results are seen for the WHU-CD dataset given in Figure \ref{fig:whu}.

    Moreover, an exhaustive ablation study was performed to demonstrate that each of the proposed components in the algorithm has a positive contribution to the result. The results are summarized in Table \ref{tab:ablation}. It can be observed that, apart from Precision, the algorithm performs best for all other metrics when all the proposed components are present. This indicates that all the elements have a meaningful contribution towards the final result.

    Additionally, the significant improvements attributed to the ECR mechanism highlight the enhanced modeling capability introduced by the depthwise separable convolution, which effectively preserves local context, unlike standard \(1 \times 1\) convolutions. Moreover, replacing conventional convolutions with depthwise separable ones for capturing spatial context leads to faster convergence and reduced computational complexity.

    Furthermore, the improvement resulting from the introduction of the DICE loss is evident and can be attributed to its effectiveness in addressing class imbalance, a prominent challenge in change detection. This benefit has also been previously demonstrated in \cite{ratnayake2025enhancedscannetcbamdice}.

    Finally, the ablation study on Channel-Wise Temporal Cross Modeling and Difference Modeling highlights their effectiveness in capturing temporal variations as well as the relationships between latent representations produced by the encoder. This is further corroborated by the sharper features observed in the change maps, as illustrated in the qualitative results in Figure~\ref{fig:combined_results}.

\section{Conclusion}
The proposed methodologies enhance remote sensing change detection through the integration of novel precision fusion blocks, an improved decoder pipeline, and a refined optimization strategy. These components facilitate the capture of fine-grained spatio-temporal changes and the optimization of key performance metrics. Experimental results demonstrate that this approach outperforms existing models in terms of accuracy and efficiency, offering a robust tool for monitoring environmental and urban transformations.

\bibliographystyle{Ref_Styles/IEEEtran}
\bibliography{references}

\end{document}